\documentclass[preprint,showpacs,nofootinbib]{revtex4}
\usepackage{graphicx}
\usepackage{epsfig}
\usepackage{dcolumn}
\usepackage{bm}
\usepackage{longtable}
\usepackage{subfigure}

\def\lsim{~\rlap{$<$}{\lower 1.0ex\hbox{$\sim$}}\;}
\def\gsim{~\rlap{$>$}{\lower 1.0ex\hbox{$\sim$}}\;}

\begin{document}

\title{Consequences of dark matter-dark energy interaction on cosmological
parameters derived from SNIa data}

\author{Luca Amendola}

\email{amendola@mporzio.astro.it}

\affiliation{INAF/Osservatorio Astronomico di Roma, V. Frascati 33, 00040, Monteporzio
Catone (Roma), Italy }

\author{Gabriela Camargo Campos}

\email{gabi@ift.unesp.br}

\author{Rogerio Rosenfeld}

\email{rosenfel@ift.unesp.br}

\affiliation{Instituto de F\'{\i}sica Te\'{o}rica - UNESP, Rua Pamplona 145 \\
01405-900, S\~{a}o Paulo, SP, Brazil}

\begin{abstract}
Models where the dark matter component of the universe interacts with
the dark energy field have been proposed as a solution to the cosmic
coincidence problem, since in the attractor regime both dark energy
and dark matter scale in the same way. In these models the mass of
the cold dark matter particles is a function of the dark energy field
responsible for the present acceleration of the universe, and different
scenarios can be parameterized by how the mass of the cold dark matter
particles evolves with time. In this article we study the impact of
a constant coupling $\delta$ between dark energy and dark matter
on the determination of a redshift dependent dark energy equation of state
$w_{DE}(z)$ and on the dark matter density today from SNIa data.
We derive an analytical expression for the luminosity distance in
this case. In particular, we show that the presence of such a coupling
increases the tension between the cosmic microwave background data
from the analysis of the shift parameter in models with constant $w_{DE}$  
and SNIa data for realistic values of the present dark matter density fraction. 
Thus, an independent measurement of the present dark matter density can place
constraints on models with interacting dark energy.
\end{abstract}


\preprint{IFT-P.035/2006}

\maketitle

\section{Introduction}

Supernovae of type Ia (SNIa) chronicle the recent expansion history
of the universe. The accumulated data encrypts information about the
composition of the universe and the physical properties of its main
components, in particular of the dark energy (DE) \cite{Reviewdarkenergy}
that drives the accelerated expansion today.

Data from SNIa \cite{SNIaRiess,SNIaLegacy}, the cosmic microwave
background (CMB) \cite{wmap} and large scale structure \cite{lss}
converged to a concordance $\Lambda$CDM model \cite{concordance},
with a nearly flat universe where a cosmological constant $\Lambda$
is the current dominant energy component, accounting for approximately
74$\%$ of the critical density, the remaining being non-relativistic
non-baryonic dark matter (DM, 22$\%$) and baryonic matter (4$\%$).

This simple ``vanilla" $\Lambda$CDM model is still compatible with data but
is not satisfactory mainly because it requires a large amount of fine
tuning in order to make the cosmological constant energy density dominant
at recent epochs.

DE can also be modelled by a scalar field, the so-called quintessence
models, either slowly rolling towards the minimum of the potential
or already trapped in this minimum 
\cite{Reviewdarkenergy,quintessence,plquintessence,expquintessence}.
In this case, the equation of state of DE may vary with cosmological
time.

In spite of the success of the concordance model, one should keep
in mind that there exists some tension between SNIa and CMB data when
the DE equation of state is allowed to be different from a cosmological
constant. Best fit models for one set of data alone is usually ruled
out by the other set at a large confidence limit \cite{tension}.
SNIa data typically favors large values of non-relativistic dark matter
abundance $\Omega_{DM}$ and a phantom-like DE equation of state $\omega<-1$.
Of course these conclusions are valid only in standard models of DE
and DM, that is, models where DE and DM are decoupled. This tension
has been ameliorated with the new data from the Supernova Legacy Survey(SNLS)
\cite{SNIaLegacy}, as shown in \cite{tension2}.

An intriguing possibility is that DM particles could interact with
the DE field, resulting in a time-dependent mass for the DM particles
and a modification in its equation of state. In this scenario, sometimes
called VAMPs (VAriable-Mass Particles) \cite{carrolvamps}, the mass
of the DM particles evolves according to some function of the dark
energy field $\phi$ such as, for example, a linear function of the field
\cite{carrolvamps,quirosvamps,peeblesvamps} with a inverse
power law dark energy potential or an exponential function 
\cite{exp1amendola,exp2amendola,exp3amendola,pietronivamps,riottovamps}
with an exponential dark energy potential. Some of these models have
a tracker solution, that is, there is a stable attractor regime where
the effective equation of state of DE mimics the effective equation
of state of DM \cite{exp1amendola,riottovamps}.

The tracker behavior is interesting because once the attractor is
reached, the ratio between DM energy density $\rho_{DM}$ and DE
energy density $\rho_{DE}$ remains constant afterwards.
This behavior could solve the ``cosmic coincidence problem'',
that is, why are the DE and DM energy densities similar today. However,
even in these cases a large amount of fine-tuning is required for the energy density 
scale of the scalar potential \cite{franca}.

In this paper we investigate the effect of a DE-DM coupling in deriving
bounds on the DE equation of state and on $\Omega_{DM}$ from SNIa
data. In particular, we show that analyzing the SNIa data with a large
positive coupling results in a larger value of $\Omega_{DM}$, thus
potentially increasing the tension with CMB data. This is confirmed by
an analysis of the CMB shift parameter \cite{shift} for interacting models 
with constant equation of state.

\section{A phenomenological model}

\label{sec:model}

Variable-mass particles generically arise in models where the quintessence
field is coupled to the non-baryonic dark matter field (coupling to
baryonic matter is severely restricted \cite{couplinglimits}). Such a coupling represents
a particularly simple and relatively general form of modified gravity:
they appear in fact in scalar-tensor models (in the Einstein frame)
and in simple versions of higher-order gravity theories in which the
action is a function of the Ricci scalar. From a lagrangian point
of view, these couplings could be of the form $g(\Phi)m_{0}\bar{\psi}\psi$
or $h(\Phi)m_{0}^{2}\phi^{2}$ for a fermionic or bosonic dark matter
represented by $\psi$ and $\phi$ respectively, where the functions
$g$ and $h$ of the quintessence field $\Phi$ can in principle be
arbitrary.

Instead of postulating a definite model by choosing two functions
defining a DE self-coupling potential and DE-DM coupling \cite{das},
one could alternatively follow an approach that is more model-independent
and closer to observations by introducing a parameterization for the
DE equation of state $w_{DE}(a)$ and for the coupling function $\delta(a)$, where $a(t)$ is the scale factor of the universe.

The dynamics of the quintessence field, governed by its potential,
induces a time dependence in the mass of dark matter particles. Therefore
one would have $m=m(\Phi(a))$ that we will parameterize in terms
of a function of the scale factor $\delta(a)$ in the following way:
\begin{equation}
m(a)=m_{0}e^{\int_{1}^{a}\delta(a')d\ln a'}\label{eq:parametrization}\end{equation}
 where $m_{0}$ is the particle mass today. In other words, in addition to
the usual parametrizations of the DE variable equation of state we
introduce a parametrization for the DM variable mass. Just as for
the equation of state $w_{DE}(a)$, this parametrization allows to study the observational data
in a systematic fashion in search of new physical phenomena.
The physical interpretation of the coupling $\delta(a)$ is therefore
straightforward: it represents the rate of change of the DM particle
mass, $\delta=d\ln m/d\ln a$.
In this paper we will focus on the simplest case, a constant coupling $\delta$.

This variable mass results in the following equation for the evolution of the DM
energy density $\rho_{DM}$:
\begin{equation}
\dot{\rho}_{DM}+3H\rho_{DM}-\delta(a)H\rho_{DM}=0\label{rhoDM}
\end{equation}
 where $H=\dot{a}/a$ is the Hubble parameter. Conservation of the
total stress-energy tensor them implies that the dark energy density
should obey
\begin{equation}
\dot{\rho}_{DE}+3H\rho_{DE}(1+w_{DE})+\delta(a)H\rho_{DM}=0.\label{rhoDE}
\end{equation}

Recently, Majerotto \textit{et al.} \cite{majerotto} (see also \cite{AGP})
considered the case of constant $w_{DE}$ and assumed a tracking behaviour
of the DM and DE densities ($\rho_{DE}/\rho_{DM}\propto a^{\xi}$)
in their analysis of SNIa data. In this class of models the modifications
of the abundance of mass-varying DM particles were studied in \cite{abundance}
and consequences for the evolution of the universe were analyzed in
\cite{CaiWang}.

However, in quintessence models the equation of state is generally
time-dependent. Therefore in this work we will study the impact of
a constant DM-DE coupling $\delta$ on the determination of a redshift
dependent DE equation of state $w_{DE}(z)$ and on the best-fit value
of the dark matter abundance today from SNIa data.

In the case of a constant interaction, eq. (\ref{rhoDM}) can be easily
solved:
\begin{equation}
\rho_{DM}(a)=\rho_{DM}^{(0)}a^{-3+\delta},\label{DMSol}
\end{equation}
 where $\rho_{DM}^{(0)}$ is the non-baryonic DM energy density today.
Substituting this solution in eq. (\ref{rhoDE}) we obtain a differential
equation in the scale factor $a$:
\begin{equation}
\frac{d\rho_{DE}}{da}+\frac{3}{a}\rho_{DE}(1+w_{DE})+\delta\rho_{DM}^{(0)}a^{-4+\delta}=0.\label{rhoDEa}
\end{equation}

Before proceeding to an evolving equation of state, it is instructive
to study the particular case of a constant $w_{DE}(a)=w$, where the
solution to eq. (\ref{rhoDEa}) is given by:
\begin{equation}
\rho_{DE}(a)=\rho_{DE}^{(0)}a^{-3(1+w)}+\frac{\delta}{\delta+3w}\rho_{DM}^{(0)}\left(a^{-3(1+w)}-a^{-3+\delta}\right)\label{DESolConstantW}
\end{equation}
 where $\rho_{DE}^{(0)}$ is the DE energy density today. The first
term of the solution is the usual evolution of DE without the coupling
to DM. From this solution it is easy to see that one must require
a positive value of the coupling $\delta>0$ in order to have a consistent
positive value of $\rho_{DE}$ for earlier epochs of the universe
(it is to be noted however that negative values of $\rho_{DE}$ could
be allowed if the dark energy is in fact a manifestation of modified
gravity, see \textit{e.g.} \cite{fr}). This feature remains in the case of varying $w$ and we will consider only positive values of $\delta$ throughout
the paper.

Furthermore, if $\delta<-3w$ one has a tracking of DE and DM densities
at earlier epochs: \begin{equation}
\frac{\rho_{DM}(a)}{\rho_{DE}(a)}\longrightarrow-\frac{\delta+3w}{\delta}\end{equation}
 resulting in \begin{equation}
\Omega_{DE}(a)=\frac{\rho_{DE}(a)}{\rho_{DE}(a)+\rho_{DM}(a)}\longrightarrow-\frac{\delta}{3w}.\end{equation}
 Therefore, requiring $\Omega_{DE}<0.1$ in the past fixes $\delta<-0.3w$
in this simple case of constant $w$. It is also interesting that
one can analytically compute the scale factor $a_{tr}$ where the
transition from DM to DE occurs in this simple case: \begin{equation}
a_{tr}=\left[-1-\frac{\delta+3w}{\delta}\frac{\rho_{DE}^{(0)}}{\rho_{DM}^{(0)}}\right]^{\frac{1}{\delta+3w}}.\end{equation}
 For instance, for $w=-1$ and $\delta=1.6$ we find $a_{tr}=0.665$
(corresponding to $z_{tr}=0.5$).

For the remaining of this work, we will study an equation of
state with the commonly used parameterization:
\begin{equation}
w_{DE}(z)=w_{0}+w_{1}z ,
\end{equation}
in order to compare with other results in the literature.
However, it should be stressed that SNIa
data is not currently very sensitive to time variations in  $w_{DE}$.
In some cases, as in the comparison with the CMB data, we will use a constant
equation of state by fixing $w_1 = 0$.
 
We obtain a closed form solution for eq. (\ref{rhoDEa}) as a function
of the redshift $z$: \begin{equation}
\rho_{DE}(z)=\rho_{DE}^{NI}(z)\left[1+\Theta(z,w_{0},w_{1},\delta)\right]\end{equation}
 where \begin{equation}
\rho_{DE}^{NI}(z)=\rho_{DE}^{(0)}e^{3w_{1}z}(1+z)^{3(1+w_{0}-w_{1})}\end{equation}
 is the usual evolution of non-interacting (NI) DE density for this
parameterization of the equation of state and the correction $\Theta$
is given by: \begin{equation}
\Theta(z,w_{0},w_{1},\delta)=\delta\; e^{3w_{1}}(3w_{1})^{3(w_{0}-w_{1})+\delta}\;\frac{\rho_{DM}^{(0)}}{\rho_{DE}^{(0)}}\Gamma(-3(w_{0}-w_{1})-\delta,3w_{1},3w_{1}(1+z))\end{equation}
 where $\Gamma(a,x_{0},x_{1})$ is the generalized incomplete gamma
function: \begin{equation}
\Gamma(a,x_{0},x_{1})=\int_{x_{0}}^{x_{1}}t^{a-1}e^{-t}dt.\end{equation}

Notice that the correction $\Theta$ vanishes in the case of no interaction,
$\delta=0$, as it should be. We have also found solutions for different
parameterizations of the dark energy equation of state but will not use them in this paper.

We use the Hubble-free dimensionless luminosity distance $d_{h}$
defined in terms of the usual luminosity distance $d_{L}$ as \begin{equation}
d_{h}=\frac{H_{0}}{c}d_{L}.\end{equation}
 The Hubble-free luminosity distance in our model for the parameterization
used is given by (we are assuming a flat universe throughout the paper):
\begin{eqnarray}
d_{h}(z,\Omega_{DM}^{(0)},w_{0},w_{1},\delta) &  & =(1+z)\int_{0}^{z}dz'\left[\Omega_{b}^{(0)}(1+z')^{3}+\Omega_{DM}^{(0)}(1+z')^{3-\delta}+\right.\nonumber \\
 &  & \left.(1-\Omega_{b}^{(0)}-\Omega_{DM}^{(0)})(1+z')^{3(1+w_{0}-w_{1})}e^{3w_{1}z'}\left(1+\Theta(z,w_{0},w_{1},\delta)\right)\right]^{-1/2}\end{eqnarray}
 where $\Omega_{DM}^{(0)}$ and $\Omega_{b}^{(0)}$ are the dark matter
and the baryonic density fractions today, respectively. In the next
section we will compare $d_{h}$ (or $d_{L}$ for a given value of
the Hubble parameter) obtained from our model to observations in order
to study the consequences of the coupling between DE and DM.

\section{Results}

\label{sec:results}

We will work with two data sets, the so-called gold data set consisting
of 157 SNIa \cite{SNIaRiess} and the recent SNLS data set of 71 new
SNIa with high redshift \cite{SNIaLegacy}. Since these two data sets
were obtained with different detection techniques and use different
methods to extract the distance moduli of the supernovae, we will
analyze them separately. Other recent analysis can be found in \cite{tension2,snls1,snls2,snls3}.

In both cases we used the maximum likelihood method for estimating
the cosmological parameters. We find the best-fit values by numerically
minimizing the likelihood function and the 68\% confidence level contour
plots were obtained by direct integration of the likelihood function.
We fixed $\Omega_{b}^{(0)}=0.05$ throughout our analysis. Since the
impact of $\Omega_{b}^{(0)}$ on $d_{h}$ is very weak, the exact
value of $\Omega_{b}^{(0)}$ is not important.

The apparent luminosity $m$ of a supernova in terms of its absolute
magnitude $M$ is usually written as:
\begin{equation}
m=M+5\log_{10}d_{h}+5\log{c H_{0}^{-1}} + 25.
\end{equation}
 The absolute magnitude of each SNIa has a spread around a standard
value: \begin{equation}
M=M_{0}+\Delta\label{delta}\end{equation}
 where $M_{0}$ is the absolute magnitude of a standard SNIa and $\Delta$
represents the corrections arising from fits to the color and light
curves of each supernova. Hence one can write the apparent magnitude
as: \begin{equation}
m_{corr}\equiv m-\Delta=5\log{d_{h}}+\bar{M}\label{aparente}\end{equation}
 where the so-called nuisance parameter $\bar{M}$ is given by \begin{equation}
\bar{M}(H_{0},M_{0})=M_{0}+5\log_{10}cH_{0}^{-1}+25.\end{equation}
 We will use the values of the distance modulus $\mu$ given by \begin{equation}
\mu=m-\Delta-\bar{M}=5\log{d_{h}}\end{equation}
 in order to derive bounds in the parameters in our model.

%
\begin{figure}[htb]
\vspace{0.5cm}
\includegraphics[scale=0.97, angle=0]{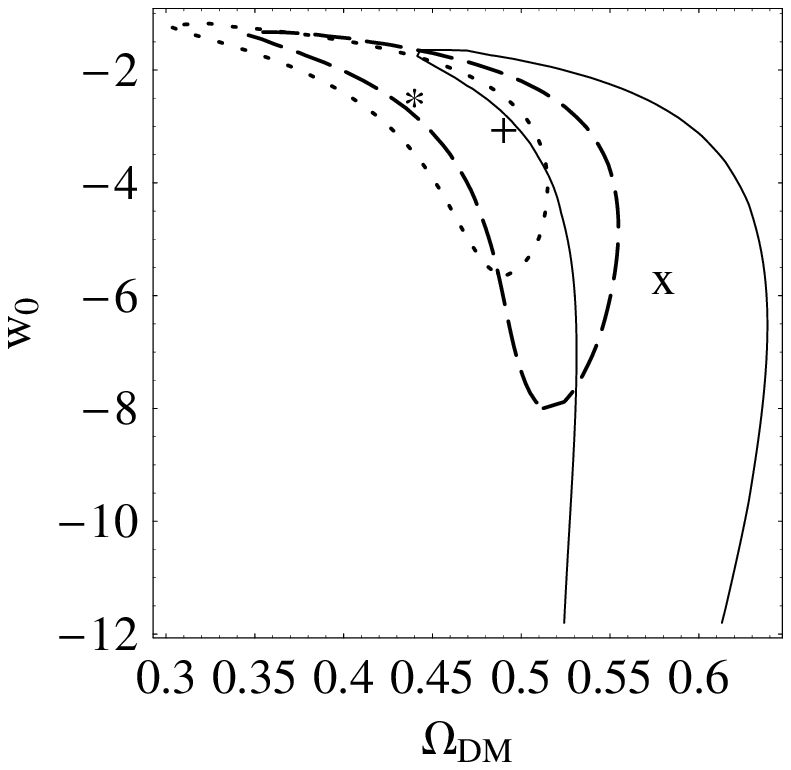}
\includegraphics[scale=1.0,angle=0]{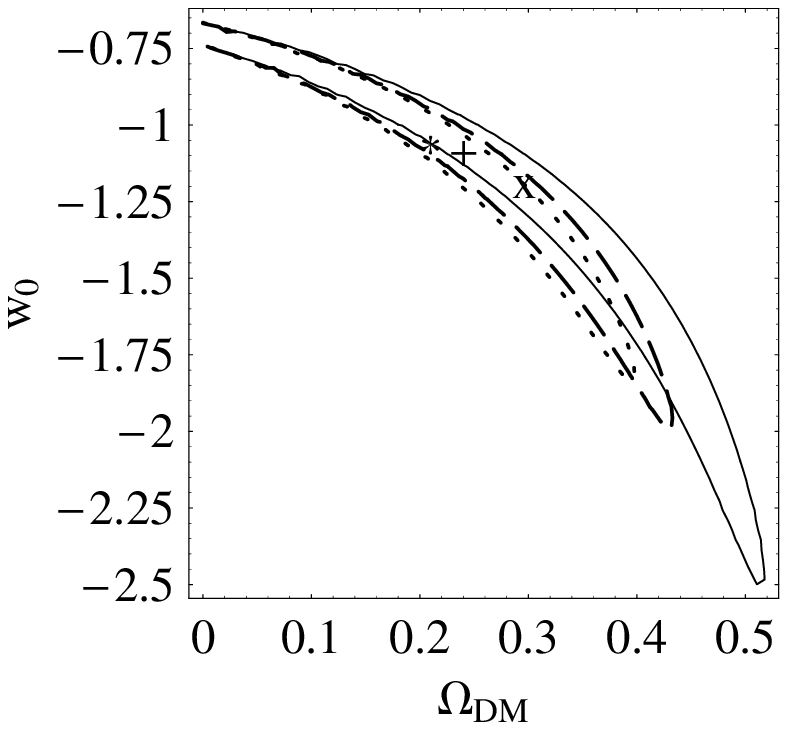}
\caption{\label{todosw10} Countour plots in the $w_{0}-\Omega_{DM}$ plane
for a constant equation of state with 68\% confidence level for different
values of the DM-DE coupling $\delta$ ($\delta=0$ (dotted line),
$\delta=0.2$ (dashed line), and $\delta=0.6$ (solid line)). The
best fit values are marked by an {}``$\ast$\char`\"{} ($\delta=0$),
{}``+\char`\"{} ($\delta=0.2$) and an {}``X\char`\"{} ($\delta=0.6$),
using the gold SNIa (left figure) and SNLS (right figure) data. }
\end{figure}

For the analysis of the SNIa gold data set, we follow ref. \cite{peri,peri2,lazkoz}
and estimate the likelihood function already marginalized over the
nuisance parameter $\bar{M}$, which includes $H_{0}$, for different
values of the coupling $\delta$.

For the analysis of the 71 new SNIa data obtained by SNLS we used
the distance moduli values $\mu_{i}$ obtained from the best fit values
(for $h=0.7$) of the correction parameters ($\alpha$ and $\beta$)
and $M_{0}$, and the corresponding errors reported in \cite{SNIaLegacy}.
The likelihood function is computed as:
\begin{equation}
{\cal L}(w_{0},w_{1},\Omega_{DM})=\exp\left[-\frac{1}{2}\sum_{i=1}^{71}\frac{(\mu_{i}-5\log d_{L}(h=0.7,w_{0},w_{1},\Omega_{DM},z_{i})-25)^{2}}{\sigma_{\mu_{i}}^{2}+\sigma_{v}^{2}+\sigma_{int}^{2}}\right]
\end{equation}
where $\sigma_{\mu_{i}}$ is the uncertainty associated with the
observational techniques in determining the magnitudes, $\sigma_{v}$
is associated with the peculiar velocities (and hence negligible for
large redshifts) and $\sigma_{int}$ is due to the intrinsic dispersion
of the absolute magnitudes.

In the case where there is no interaction between DE and DM, we find
that the best fit values for $\Omega_{DM}^{(0)}$ in a $\Lambda$CDM
model ($w_{1}=0$ and $w_{0}=-1$) are $\Omega_{DM}^{(0)}=0.26\pm0.04$
and $\Omega_{DM}^{(0)}=0.19\pm0.02$ for the gold set and SNLS data
respectively. The best fit from the SNLS data is in remarkable agreement
with the recent analysis of the 3-year data from the Wilkinson Microwave
Anisotropy Probe (WMAP3y) \cite{wmap} alone, which results in 
$\Omega_{DM}^{(0)}=0.18\pm0.04$,
whereas the best fit from the gold data set has a somewhat poorer
agreement.

Many parameterizations for the DE equation of state were tested with
SNIa data but frequently fixing a particular value of $\Omega_{M}^{(0)}$
or marginalizing over a flat or gaussian prior around $\Omega_{M}^{(0)}=0.27$
\cite{SNIaRiess,tension,peri,peri2,lazkoz}. However, the agreement between SNIa and CMB gets particularly worse for the gold data set when we allow for $w_{DE}(z)$ and $\Omega_{DM}^{(0)}$ to vary simultaneously without any priors.
For instance, fixing $w_{1}=0$, that is, a constant DE equation of
state, we find the best fit values are $w_{0}=-2.4$ and $\Omega_{DM}^{(0)}=0.44$
for the gold set and $w_{0}=-1.0$ and $\Omega_{DM}^{(0)}=0.21$ for
the SNLS data. Allowing for an evolving equation of state does not
significantly alter the fits. A possible tension between SNIa and
CMB data which was present in the gold data set when one considers
models other than the $\Lambda$CDM model practically disappeared
in the SNLS data \cite{tension2}.

We want to investigate in this article the effects of adding a coupling
between DE and DM on these fits to SNIa data. The coupling will be
characterized by the constant $\delta$. We analyze first the case of constant
$w$ (i.e. $w_1=0$) and then generalize to a variable equation-of-state.

\subsection{ Constant equation of state}
We show in Figure \ref{todosw10} the 68\% confidence level contour
plots in the $w_{0}\times\Omega_{DM}^{(0)}$ plane for different values
of the coupling $\delta$, keeping $w_{1}=0$ (constant equation of
state). Notice that the SNLS data results in a better agreement with
CMB measurements. One can see the existence of a correlation between
$w_{0}$ and $\Omega_{DM}^{(0)}$ and that turning on the interaction
results in a marked tendency towards increasing the best value for
$\Omega_{DM}^{(0)}$, against the CMB results.

However, a direct comparison
of our result for $\Omega_{DM}^{(0)}$ with the CMB results is not possible,
since the latter has been obtained in the context of uncoupled models.
It is therefore important to see whether there are upper limits to
$\Omega_{DM}^{(0)}$ which are independent of the cosmological model. An
upper limit to $\Omega_{DM}^{(0)}$ which does not depend on the background
cosmology can be obtained from the galaxy cluster dynamics. However
the current data yield very weak constraints: ref.~\cite{Feldmanetal}
gives $\Omega_{m}^{(0)}=0.30_{-0.07}^{+0.17}$
so that even $\Omega_{DM}^{(0)}=0.6$ is not excluded at more than
95\% c.l. and actually the strong degeneracy with $\sigma_{8}$ allows
for even higher values.

%
\begin{figure}[htb]
  \begin{center}
    \mbox{
       \subfigure{\includegraphics[scale=0.8, angle=0]{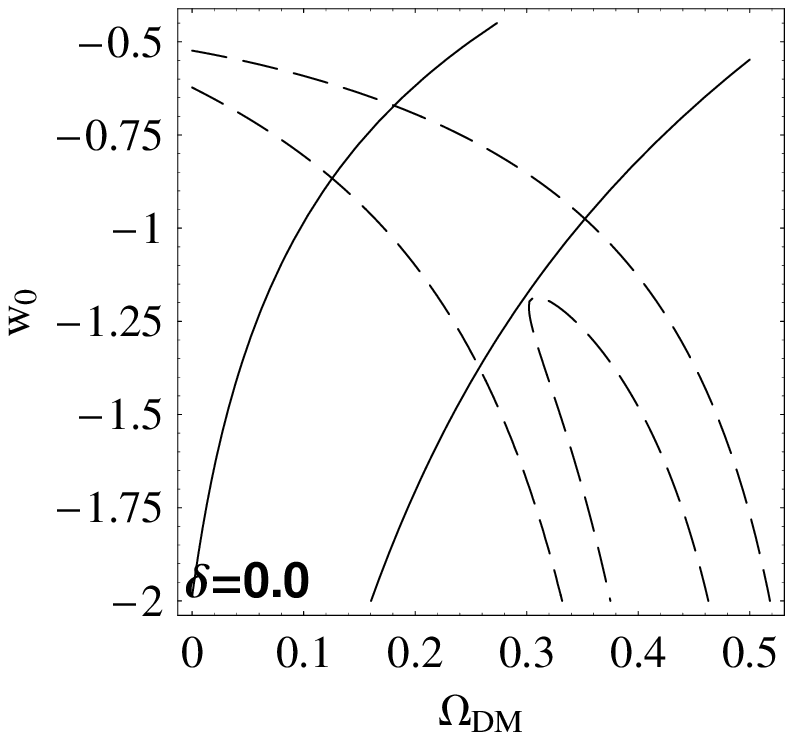}} \quad    \subfigure{\includegraphics[scale=0.8,angle=0]{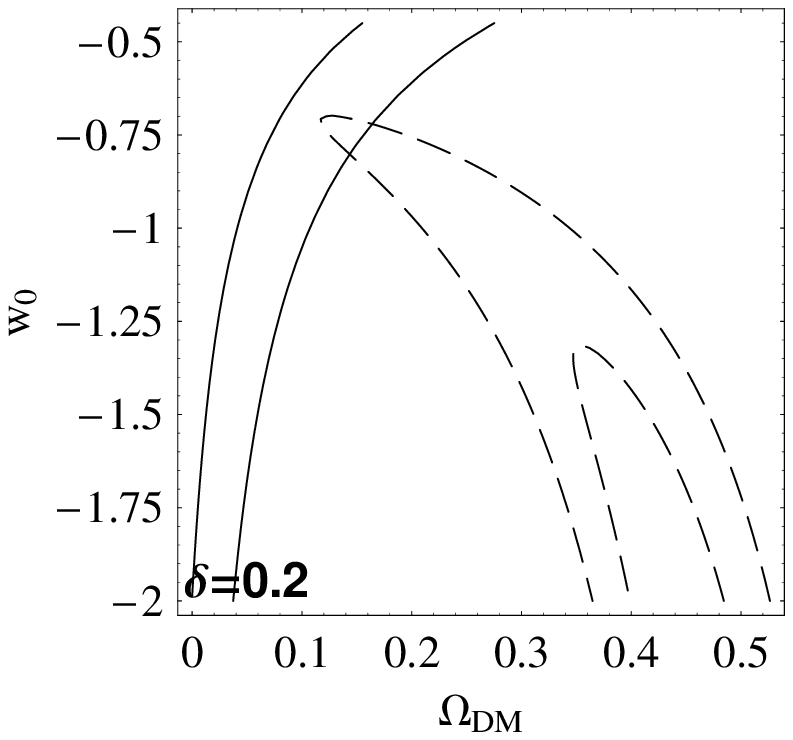}}
      }
    \mbox{   \subfigure{\includegraphics[scale=0.8,angle=0]{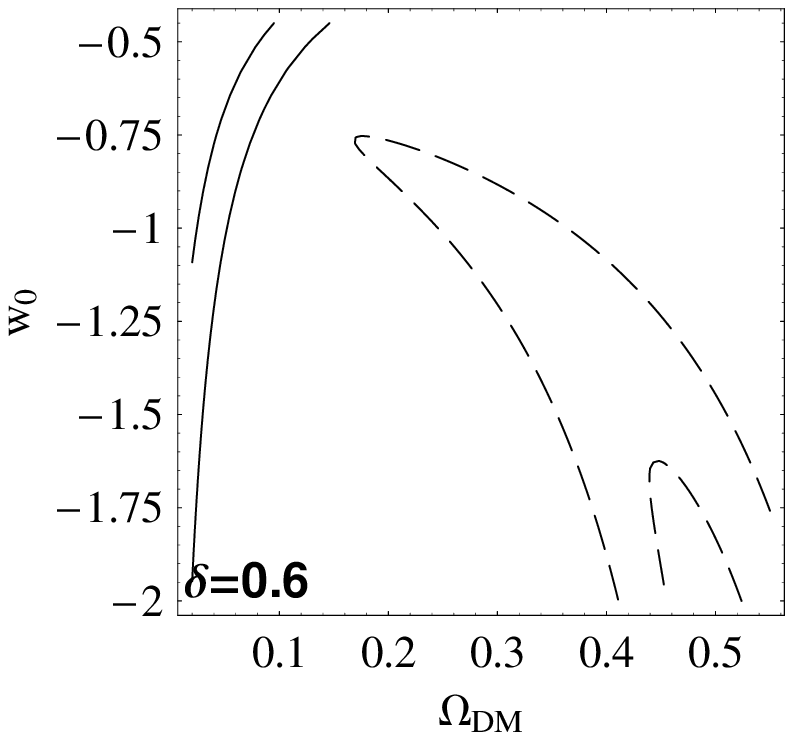}
 } \quad    \subfigure{\includegraphics[scale=0.8,angle=0]{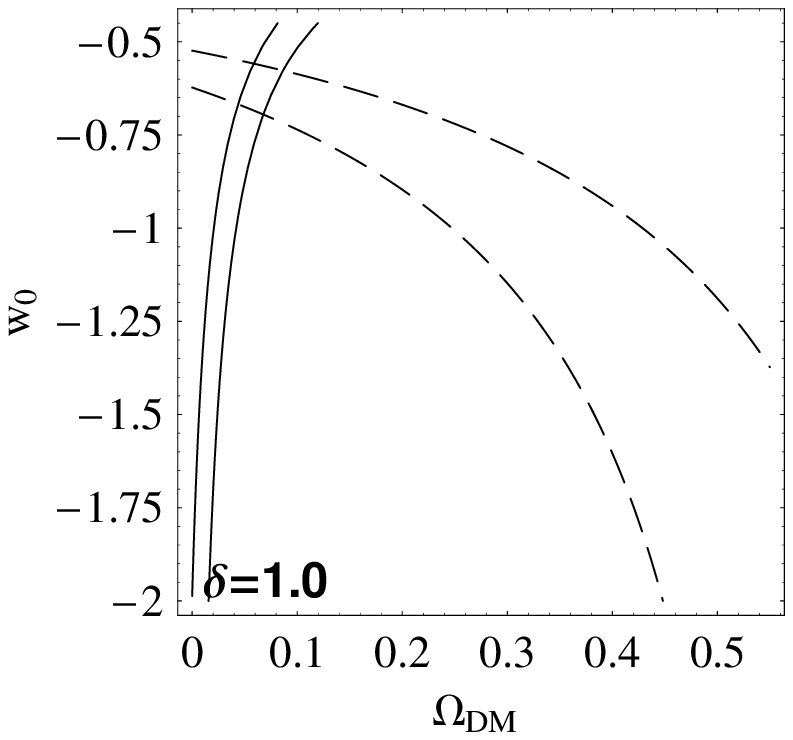}}
      }
    \caption{Countour plots in the $w_{0}-\Omega_{DM}$ plane
for a constant equation of state with $3 \sigma$ confidence level for different
values of the DM-DE coupling $\delta$. We compare constraints from
the gold data set SNIa (solid line) and CMB (dashed line) data.}
    \label{CMBRiess}
\end{center}
\end{figure}

%
\begin{figure}[htb]
  \begin{center}
    \mbox{
      \subfigure{\includegraphics[scale=0.8, angle=0]{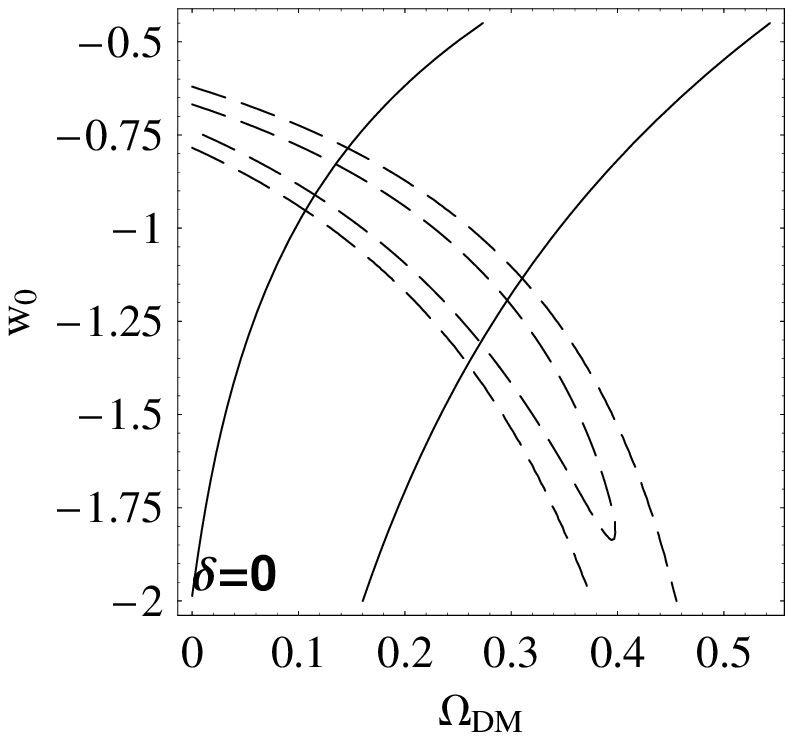}} \quad      \subfigure{\includegraphics[scale=0.8,angle=0]{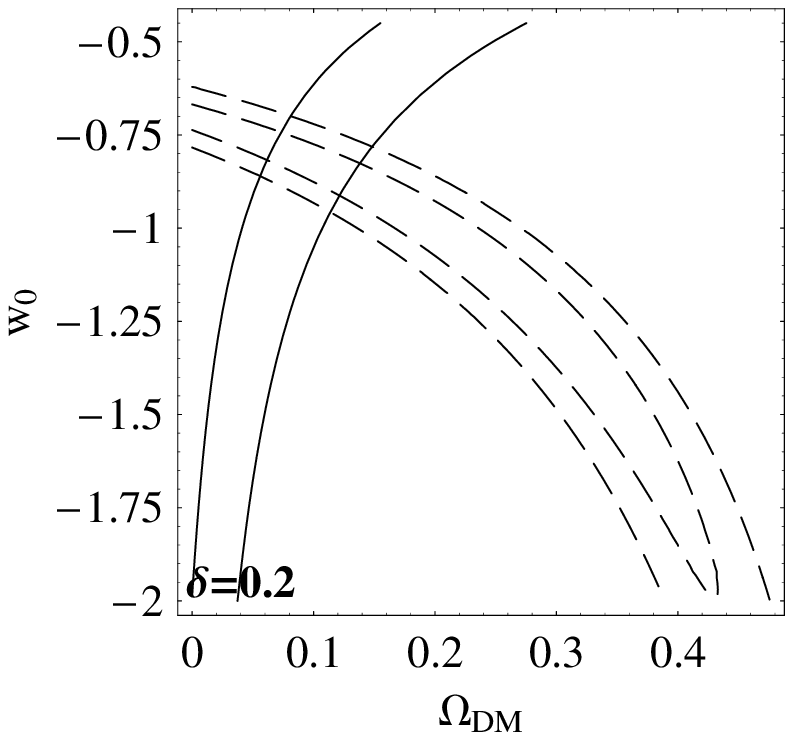}}
      }
    \mbox{      \subfigure{\includegraphics[scale=0.8,angle=0]{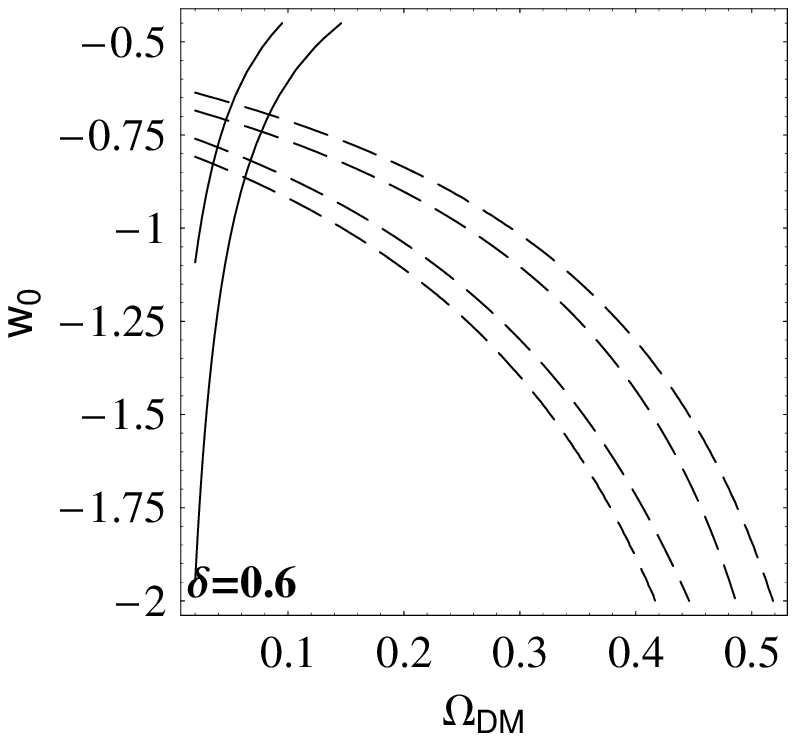}
 } \quad      \subfigure{\includegraphics[scale=0.8,angle=0]{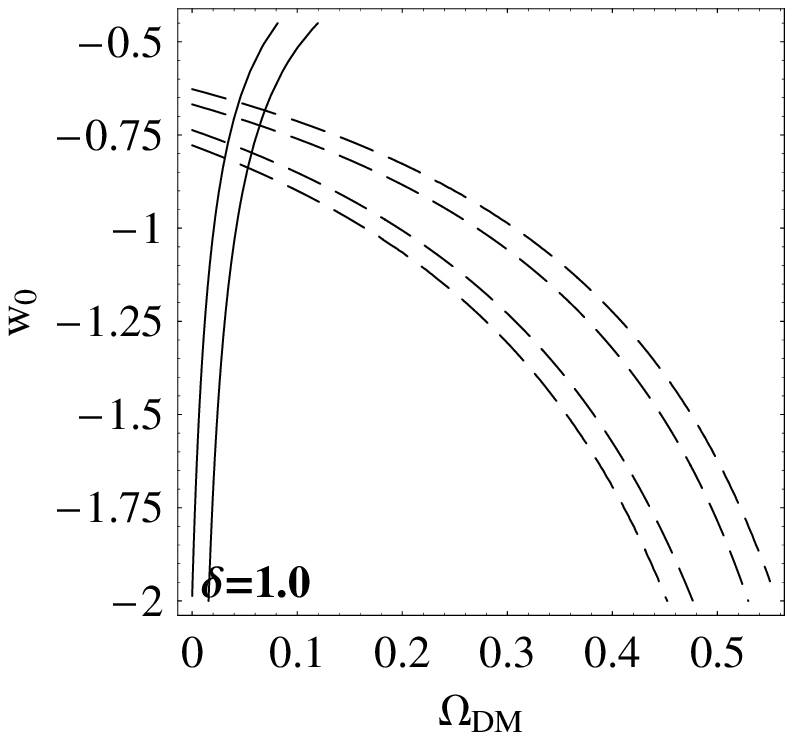}}
      }
\caption{\label{CMBSNLS} Countour plots in the $w_{0}-\Omega_{DM}$ plane
for a constant equation of state with $3 \sigma$ confidence level for different
values of the DM-DE coupling $\delta$. We compare constraints from
the SNLS SNIa (solid line) and CMB (dashed line) data. }
\end{center}
\end{figure}

On the other hand, for a constant equation of state, we can use
the CMB shift parameter in order to study the effects of interaction.
The shift parameter encapsulates information contained in the detailed 
CMB power spectrum and, more importantly, its measured values is weakly
dependent of assumptions made about dark energy. It has been used recently to
put constraints on models with braneworld cosmologies \cite{usedshift}.

The shift parameter $R$ is defined by:
\begin{equation}
R = \Omega_{M}\int_{0}^{z_{dec}}\frac{dz'}{E(z')}
\end{equation}
where $z_{dec} = 1088$ is the redshift of matter-radiation decoupling
and $E(z)=H(z)/H_{0}$. We will use the value $R=1.70 \pm 0.03$, obtained from
WMAP3y data \cite{shift}. Clearly for such calculations a radiation component with 
standard conservation equation was appropriately included.

In Figs.~\ref{CMBRiess} and \ref{CMBSNLS} we compare the resulting 
$3 \sigma$ bounds obtained from $R$ and from SNIa data. 
Figure \ref{CMBRiess} clearly shows the tension between CMB and the gold 
data set referred to earlier. It is clear that the introduction of the coupling
makes the CMB and the gold data set more incompatible with each other.
The coupling favors smaller values of $\Omega_{DM}$ from CMB at the same time
favoring larger values of the same quantity from SNIa data.
The same qualitative behaviour is seen for the SNLS data, although since there is already 
a good agreement with $\delta=0$, the situation is less drastic in this case.
It is interesting to notice that assuming a constant $w_{DE} \simeq -1$ and requiring
$\Omega_{DM} \gsim 0.1$ seems to rule out a coupling $\delta \gsim 0.5$ from 
the bounds arising from the shift parameter.

\subsection{ Variable  equation of state}
\begin{figure}[htb]
\vspace{0.5cm}
 \includegraphics[scale=0.41, angle=0]{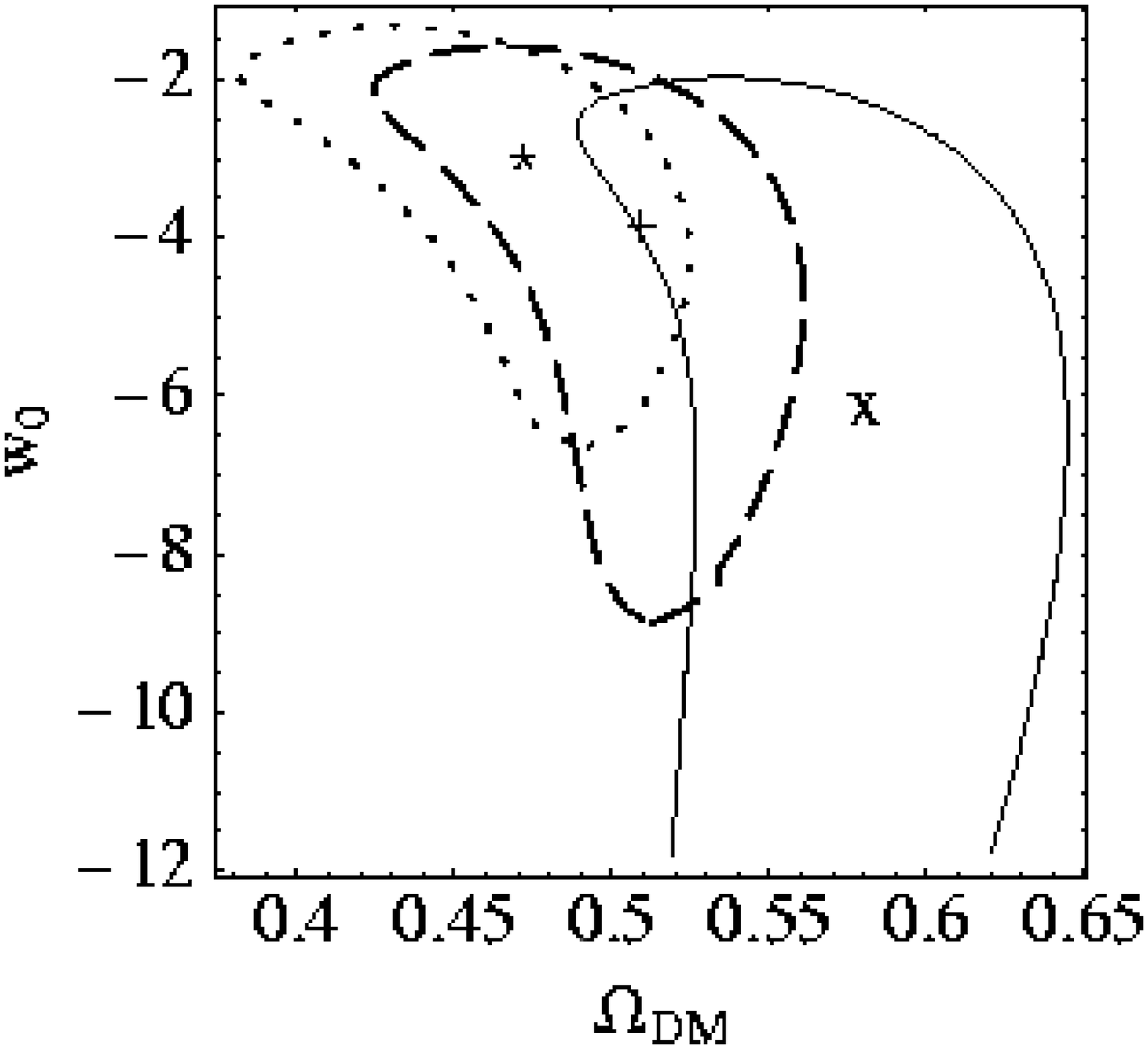}
\includegraphics[scale=0.98, angle=0]{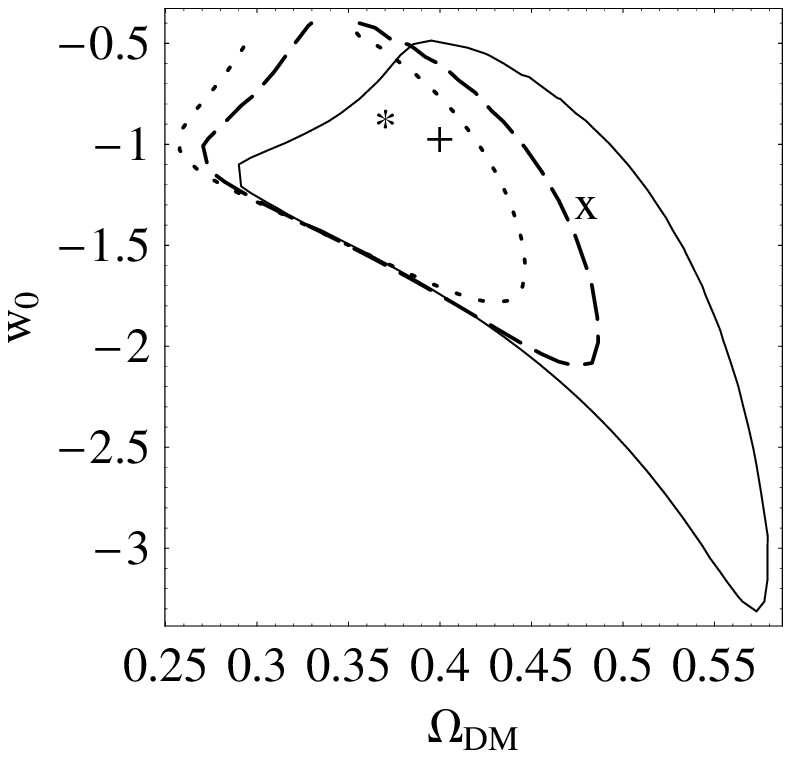}
\caption{\label{omw0} Countour plots in the $w_{0}-\Omega_{DM}$ plane marginalized
over values of $w_{1}$ with 68\% confidence level for different values
of the DM-DE coupling $\delta$ ($\delta=0$ (dotted line), $\delta=0.2$
(dashed line), and $\delta=0.6$ (solid line)). The best fit values
are marked by an {}``$\ast$\char`\"{} ($\delta=0$), {}``+\char`\"{}
($\delta=0.2$) and an {}``X\char`\"{} ($\delta=0.6$), using the
gold SNIa (left figure) and SNLS (right figure) data. }
\end{figure}

Allowing $w_{1}\neq0$ does not change the qualitative features of
Figure \ref{todosw10}. We show in Figure \ref{omw0} the $68\%$
confidence level contour plots in the $w_{0}\times\Omega_{DM}^{(0)}$
plane obtained with a marginalization over $w_{1}$ and as expected
the allowed region gets somewhat broader, with the best fit of $\Omega_{DM}^{(0)}$
shifting to larger values.

%

\begin{figure}[htb]
\vspace{0.5cm}
\includegraphics[scale=1.0,angle=0]{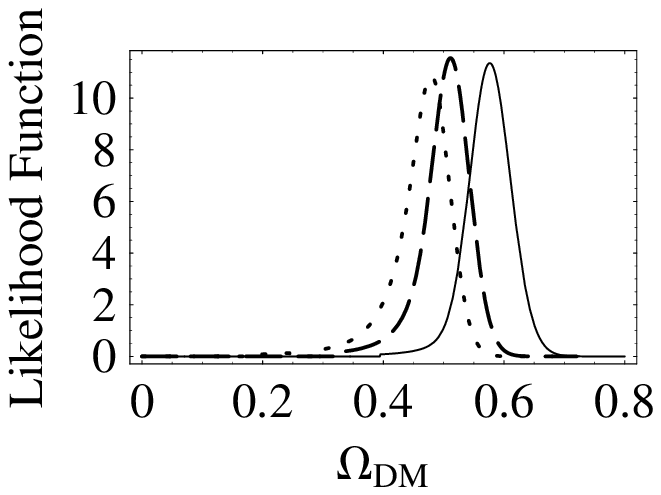}
\includegraphics[scale=1.0,angle=0]{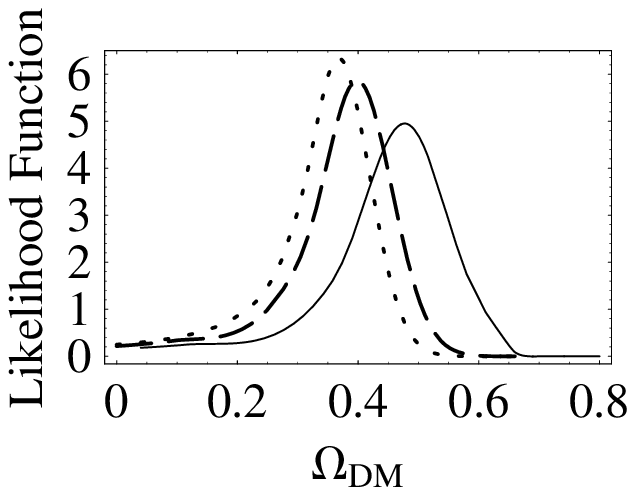}
\caption{\label{like} Plot of the likelihood function for $\Omega_{DM}$
with $w_{0}$ and $w_{1}$ marginalized for different values of the
DM-DE coupling $\delta$ ($\delta=0$ (dotted line), $\delta=0.2$
(dashed line), and $\delta=0.6$ (solid line)) using the gold SNIa
(left figure) and SNLS (right figure) data. }
\end{figure}

In Figure \ref{like} we show the marginalized likelihood function
over $w_{0}$ and $w_{1}$ in order to obtain the $68\%$ confidence
level estimation for $\Omega_{m}$. The main effect of the coupling
is to increase the estimation of $\Omega_{DM}^{(0)}$. It is interesting
to observe that for the SNLS dataset the WMAP3y value for $\Omega_{DM}$
is rejected at more than 95$\%$ c.l. already for $\delta\ge0.2$:
however, as we anticipated, such a direct comparison is not correct.

These results are summarized in Table I, where we show the best fit
values of $w_{0}$ and $\Omega_{DM}$ with their corresponding $1\sigma$
errors obtained by marginalizing over $w_{1}$ and $\Omega_{DM}$
(in the case of $w_{0}$) and $w_{1}$ and $w_{0}$ (in the case of
$\Omega_{DM}$) for both sets of SNIa data. Although the best fit values
for $\Omega_{DM}$ are quite large when $\delta$ is included, the
distribution is non-zero even for small values of $\Omega_{DM}$.

\begin{table}
\vspace{10pt}
\begin{center}
\begin{tabular}{|r|r|r|r|r|}
\hline
  &Gold&SNLS&Gold&SNLS\\
\hline
$\delta$&$w_0$&$w_0$&$\Omega_{DM}$&$\Omega_{DM}$\\
\hline
0.0&$-2.6^{+1.2}_{-2.1}$&$-0.95^{+0.32}_{-0.42}$&$0.48^{+0.03}_{-0.04}$&$0.37^{+0.06}_{-0.08}$\\
0.2&$-3.2^{+1.5}_{-2.8}$&$-1.00^{+0.35}_{-0.52}$&$0.51^{+0.04}_{-0.05}$&$0.40^{+0.07}_{-0.08}$\\
0.6&$-6.0^{+2.4}_{-3.7}$&$-1.15^{+0.42}_{-0.82}$&$0.58^{+0.03}_{-0.04}$&$0.48^{+0.08}_{-0.09}$\\
\hline
\end{tabular}
\end{center}
\caption{ Best fits obtained for $w_0$ and
$\Omega_{DM}$ for both sets of SNIa data.}
\end{table}

\begin{figure}[htb]
\vspace{0.5cm}
  \includegraphics[scale=0.96, angle=0]{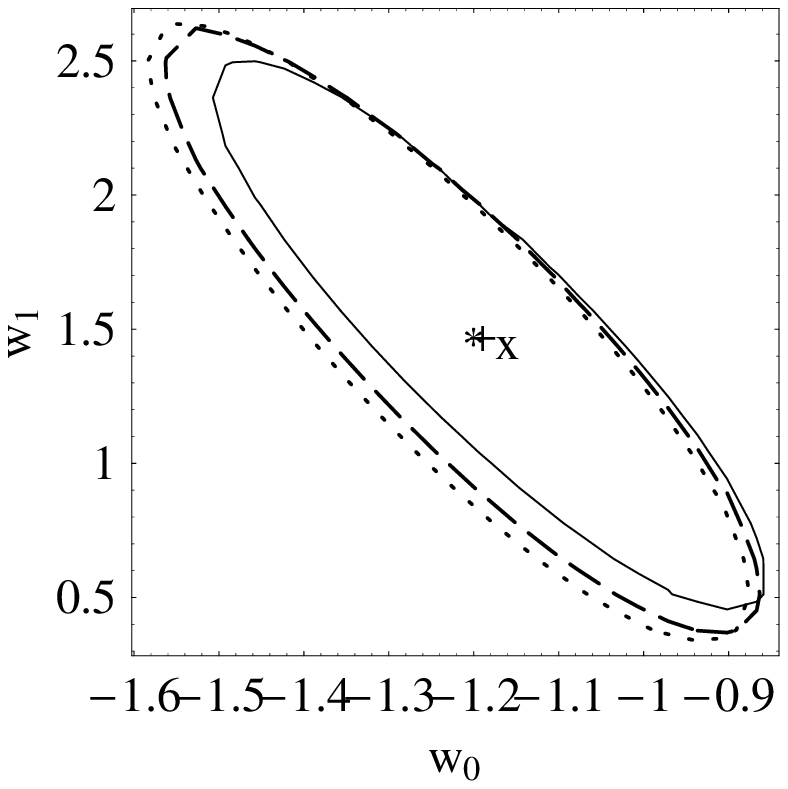}
\includegraphics[scale=0.96, angle=0]{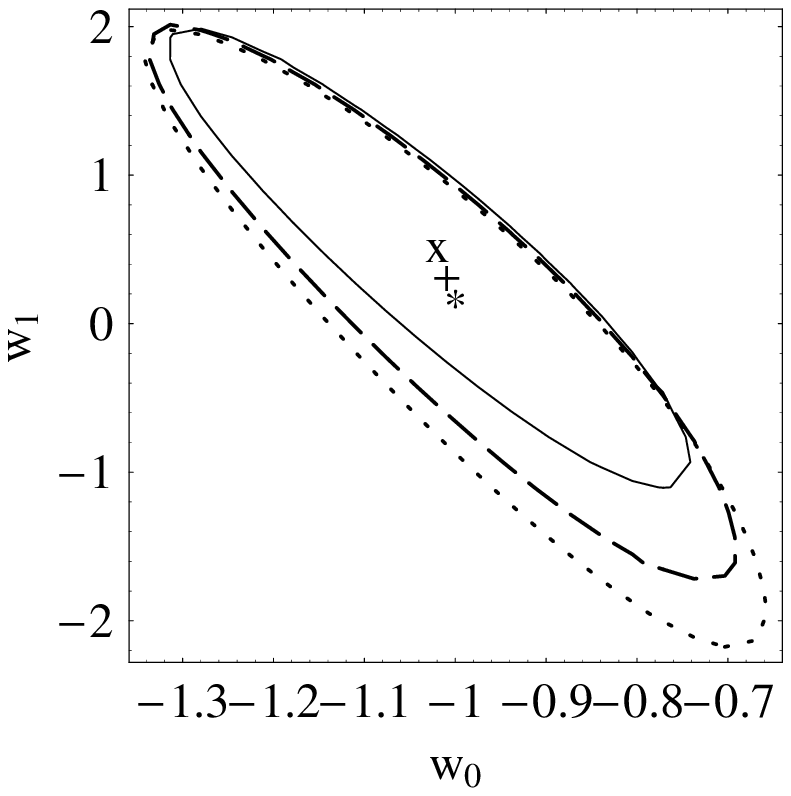}
\caption{\label{w0w1} Countour plots in the $w_{0}-w_{1}$ plane marginalized
with a gaussian prior over values of $\Omega_{DM}$ with 68\% confidence
level for different values of the DM-DE coupling $\delta$ ($\delta=0$
(dotted line), $\delta=0.2$ (dashed line), and $\delta=0.6$ (solid
line)). The best fit values are marked by an {}``$\ast$\char`\"{}
($\delta=0$), {}``+\char`\"{} ($\delta=0.2$) and an {}``X\char`\"{}
($\delta=0.6$), using the gold SNIa (left figure) and SNLS (right
figure) data. }
\end{figure}

It is also interesting to study the consequences of the DM-DE coupling
in the determination of the parameters characterizing the DE equation
of state. In Figure \ref{w0w1} we plot the $68\%$ confidence level
contour plots in the $w_{0}\times w_{1}$ plane obtained by marginalizing
the likelihood function over $\Omega_{DM}$ using a gaussian prior
$\Omega_{DM}=0.18\pm0.04$, as obtained from the WMAP3y data. The effect
of the coupling is very small when the prior on the DM density is
taken into account.

\section{Conclusions}

There is a vast amount of work studying the possibility of having
an interaction between the dark energy and the dark matter components
of our universe. In this paper we analyzed a simple model for the
interaction between these two fluids, in which the mass of the dark
matter particles increases at a constant rate. We have shown that
introducing a coupling between the dark energy component of the universe
with dark matter particles has the effect of increasing the best fit
values of the DM density today obtained from current SNIa data.

We performed a comparison with CMB data for the case of constant equation of state, using
the shift parameter. We found that the introduction of the coupling results in a poorer 
compatibility between CMB and SNIa data.
Our results showed that assuming a constant $w_{DE} \simeq -1$ and requiring
$\Omega_{DM} \gsim 0.1$, a coupling  $\delta \gsim 0.5$ seems to be ruled out.
This must be checked by model-independent measurements of $\Omega_{DM}$ from large scale 
structure and also with a more careful analysis of the CMB data including more observables 
in addition to the shift parameter.
We also found that introducing the coupling does not change significantly the 
determination of the DE equation of state when a prior on $\Omega_{DM}$ is adopted.

We worked with a simple parameterization of the dark energy equation
of state and for the DM-DE coupling but we believe that our results are 
fairly general for 
the type of interaction that we introduced. It would be interesting
to extend our analysis to more general parameterizations for both the interaction and
the equation of state available in the literature. However, it should be stressed that SNIa
data is not currently very sensitive to a time-varying  $w_{DE}$.

\section*{Acknowledgments}

This work was supported by Funda\c{c}\~{a}o de Amparo \`{a} Pesquisa
do Estado de S\~{a}o Paulo (FAPESP), grant 01/11392-0, and by Conselho
Nacional de Desenvolvimento Cient\'{\i}fico e Tecnol\'{o}gico (CNPq).

\end{document}